\documentclass[journal]{IEEEtran}
\usepackage{cite}
\usepackage{longtable}

\ifCLASSINFOpdf
 \usepackage[pdftex]{graphicx}
\else
\fi
\usepackage[cmex10]{amsmath}
\usepackage{algorithmic}
\usepackage{array}
\usepackage{mdwmath}
\usepackage{mdwtab}
\usepackage{eqparbox}
\usepackage{longtable}

\usepackage[tight,footnotesize]{subfigure}
\usepackage{caption}
\usepackage[font=footnotesize]{subfig}
\usepackage{pdflscape}

\begin{document}
%
%paper title
\title{Beamforming Techniques for Large-N Aperture Arrays}

\author{K.~Zarb-Adami, A.~Faulkner, J.G.~Bij de Vaate, G.W. Kant and P.Picard
        % <-this % stops a space
\thanks{Kristian~Zarb-Adami is with the Department
of Astrophysics, University of Oxford,
email:kza@astro.ox.ac.uk}% <-this % stops a space
\thanks{A. Faulkner is at the University of Cambridge, Cavendish Laboratory, Cambridge CB3 0HE, UK}
\thanks{J.G. de Vaate and G.W. Kant are at ASTRON, Netherlands}
\thanks{P. Picard is at the Observatoire de Paris at Nancay}
% <-this % stops a space
\thanks{Manuscript submitted June 30, 2010}}

\maketitle

\begin{abstract}
%\boldmath
Beamforming is central to the processing function of all
phased arrays and becomes particularly challenging with a large number of
antenna element (e.g. $>$100,000). The ability to beamform efficiently with reasonable power requirements is discussed in this paper. Whilst the most appropriate beamforming technology will change over time due to semiconductor and processing developments, we present a hierarchical structure which is technology agnostic and describe both Radio-Frequency (RF) and digital hierarchical beamforming approaches. We present implementations of both RF and digital beamforming systems on two antenna array demonstrators, namely the Electronic Multi Beam Radio Astronomy ConcEpt (EMBRACE) and the dual-polarisation all-digital array (2-PAD). This paper will compare and contrast both digital and analogue implementations without considering the deep system design of these arrays.
\end{abstract}



% \fi
\IEEEpeerreviewmaketitle

\section{Introduction}
\begin{figure}
\begin{center}
\includegraphics[width=2.5in]{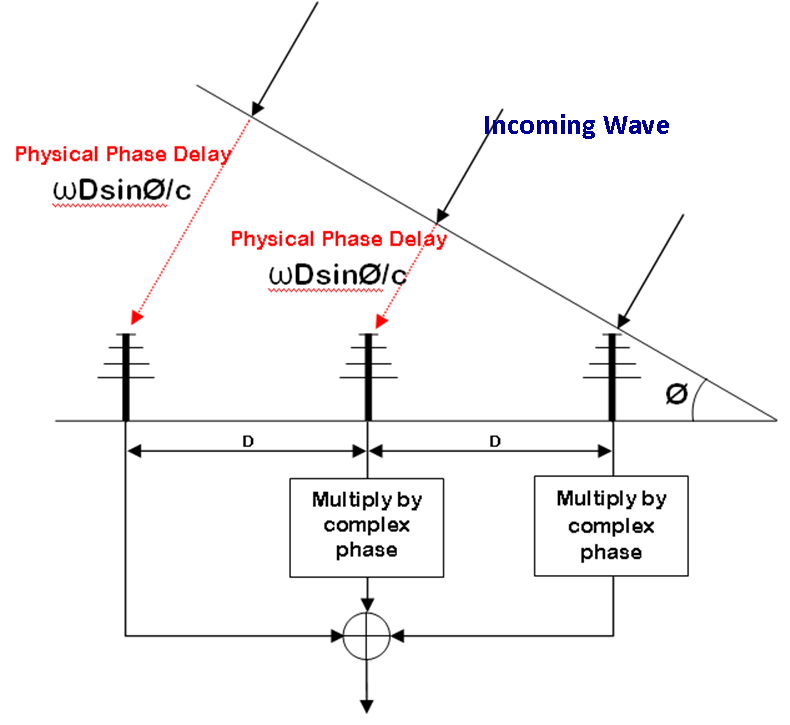}
 \caption{The Beamforming operation.}
\label{fig_overall_bf}
\end{center}
\end{figure}

\IEEEPARstart{T}{he} purpose of beamforming is to precisely align the phases of an incoming signal from different parts of an array to form a well understood beam in a specific direction. This is illustrated in Figure \ref{fig_overall_bf} for one dimension. Essentially, the signals from each of the elements are delayed such that when they are summed they all have the same delay corresponding to a specific direction. There is a physical geometric delay on the incoming wavefront that increases linearly across the array, which is then compensated for
electronically prior to being summed. This approach will work perfectly across a wide frequency
bandwidth if the signals are delayed in time and then summed (within the limits of the antenna and receiving system electronics).

The issue with this straightforward concept is that, particularly for RF/analogue beamforming, a true time delay is relatively bulky and expensive to implement, typically being done using extended, switchable signal path lengths often with lengths of circuit board tracks or cables. These often suffer from environmental and repeatability issues. The alternative is to approximate the time delay with phase delays, which is considerably easier to implement electronically particularly when many signals are being combined.

\subsection{Time-Delay Beamforming}

This is the ideal approach for beamforming since, if it can be implemented it is inherently wide bandwidth and necessary for the long delays across the array. However, the implementation is complicated in the analogue domain and few phase steps will limit resolution, which will also introduce errors in the final beam. Furthermore, to implement multiple beams with an analogue implementation simply multiplies the amount of hardware required for every beam, resulting a complex system.

In the digital domain, after digitizing the signal, it is relatively simple to introduce delays which are a multiple of the sample time and simply buffer the signals from elements which require a delay. Higher time resolution, however, is required than just the sample time. This can be achieved by interpolating between two or more samples, the effects of which need a more sound theoretical analysis. There is some complexity involved but it is eminently achievable, even for multiple beams given enough processing power. There are questions on the accuracy of the interpolation techniques and what effects that will have on the performance of the array which are described in detail in Alexander et al. [\ref{Alexander}].

%\IEEEpubidadjcol

\subsection{Phase Shift Beamforming}

Time delays can be introduced by using phase shifting systems in a narrow frequency range. This is easier to do in both the analogue and digital domains. The basic problem with using phase delays for the time delay element is that it makes the delay frequency dependant, so the time delay at the top of the band is shorter than the time delay at the bottom of the band, hence the time delay will only be accurate in the centre of the band. The bandwidth is therefore limited by the errors that can be sustained at the edges of the band. This effect can be mitigated by processing multiple narrower bands to make up the overall bandwidth required.
To introduce multiple bands in RF beamforming is expensive; in effect it is necessary to have the repeat the same amount of electronics for each additional beam. Hence, an analogue system tries to maximize the fractional beamwidth for best use of the electronics.

In a digital system the bandwidth issue can be solved by splitting the bandwidth into arbitrarily many narrow bands and then phase shifting each band for the time delay required. This is expensive computationally since each input stream needs to be put through an FFT or polyphase filter process. Phase shifting is then a straightforward complex multiplication function on each of the sub-bands. The sub-bands can be made narrow enough to make any errors negligible. There are important other benefits when the signal is split into many narrow bands: Interference filtering of narrow frequency spikes can be eliminated early in the signal chain by attenuating the relevant sub-bands; maybe more importantly for phased arrays with a large fractional bandwidth is the ability to accurately calibrate the band pass characteristics of the analogue signal chain from the element into the digitizer. This should reduce the cost of implementation of the analogue gain chain.

\section{Hierarchical Beamforming}

\begin{figure}
\begin{center}
\includegraphics[width=3.0in]{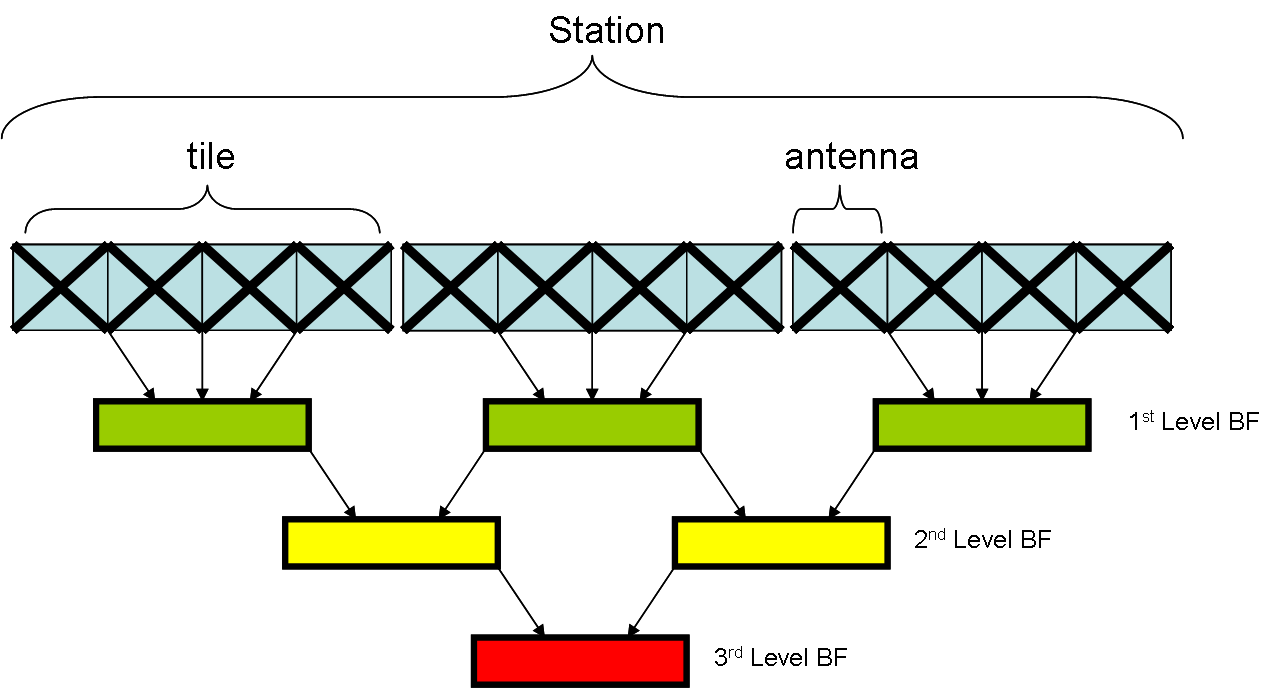}
 \caption{A Hierarchical beamforming scheme showing how antennas are grouped into `tiles' and then `stations'}
\label{fig_hier_bf}
\end{center}
\end{figure}

\begin{figure}
\begin{center}
\includegraphics[width=3.0in]{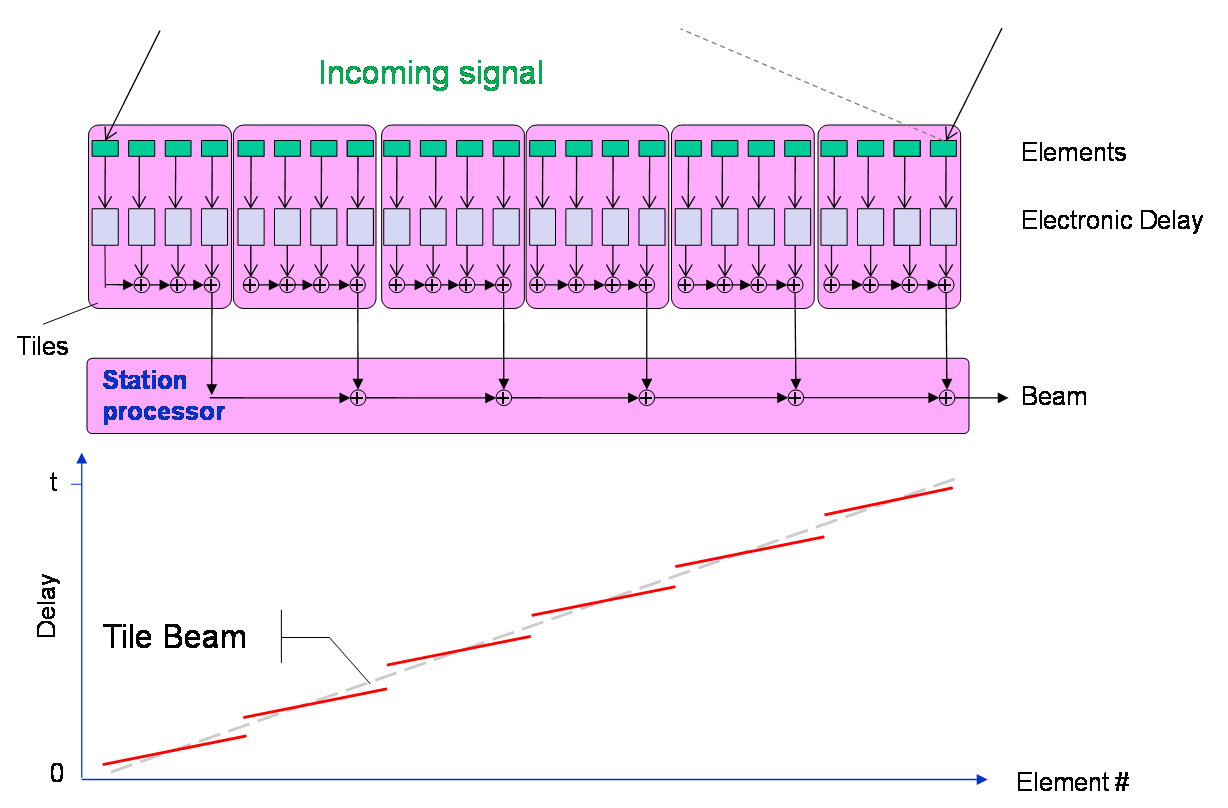}
 \caption{Figure showing how hierarchical beamforming can introduce errors in the pointing direction of `tile' beams.}
\label{bf_error}
\end{center}
\end{figure}

In order to beamform a large number of antennas the beamforming design needs to have a hierarchical structure due to the processing and the data-transport overhead. However, as with most reduction techniques there will be a level of compromise involved. By processing elements in groups (the size of which is determined by either the processing or the communications requirements), as `tiles', limited numbers of tile beams are used to create a larger number of `station' beams. This structure shown in figure \ref{fig_hier_bf} leads to errors in the beamforming which could increase the further off-centre from a tile beam it is. The way these errors manifest themselves is shown in figure \ref{bf_error} and described in more detail in [\ref{Alexander}].

\section{RF Beamforming}

As has been discussed, in a practical implementation of a large phased array beamforming is a hierarchical process and different techniques and technologies can be used at different levels of the hierarchy. The high volume of the beamforming work is performed at the `tile level' or first stage immediately after the elements. It is generally accepted that `station level' or higher level beamforming after the tiles will need to be in the digital domain for the required flexibility and field-of-view (FoV).

RF beamforming takes the analogue signals after amplification by a low noise amplifier, and implements appropriate delays through true time delay or phase shifting and then sums the results to make a beam. Each polarisation of each beam needs hardware dedicated to the task, although multiple channels can be implemented in each chip. As can be seen this is a relatively simple process, but does require careful design of the analogue electronics to ensure stability. The issues are that it is a phase shift beamformer operating with a single frequency channel, which limits available bandwidth and the resolution of the phase shifter and gain control is limited.

\begin{figure}
\begin{center}
\includegraphics[width=2.5in]{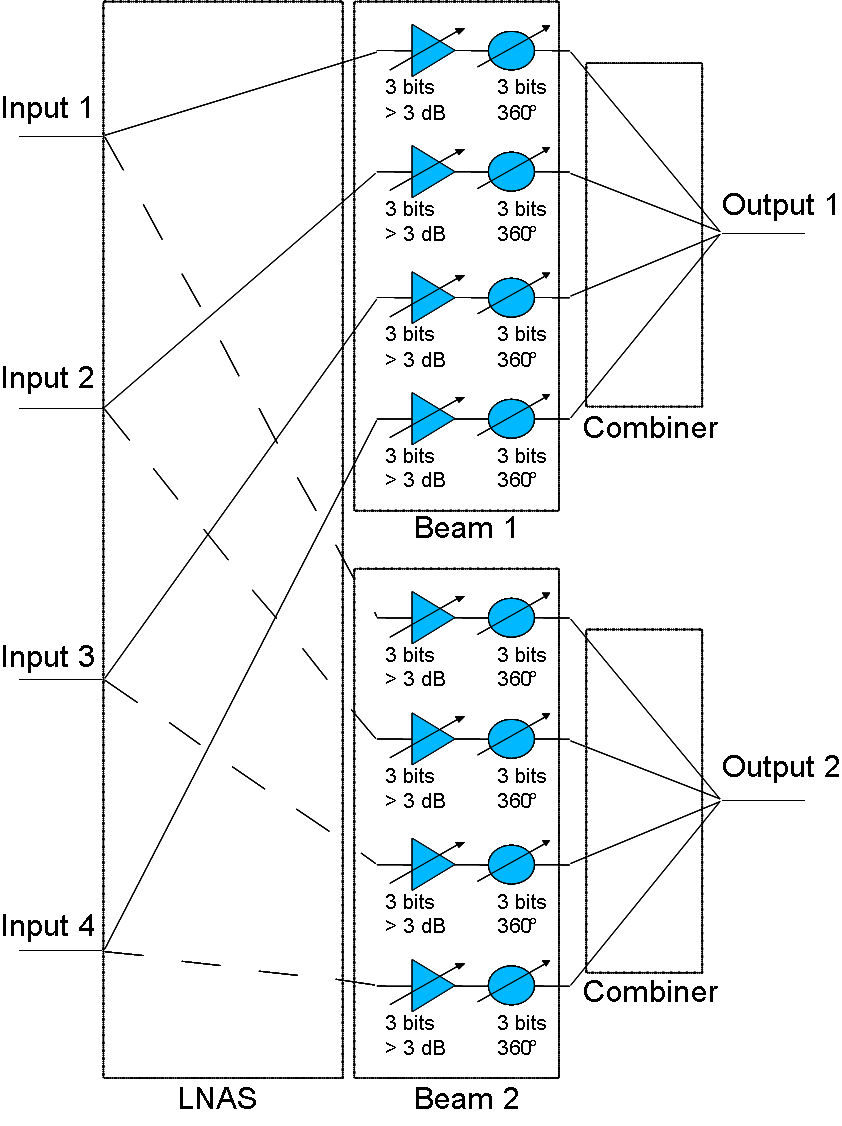}
 \caption{The Beamforming chip designed for the EMBRACE system.}
\label{fig_embrace_bf}
\end{center}
\end{figure}

The block diagram of an analogue beamformer chip implemented on the EMBRACE system is shown in Figure \ref{fig_embrace_bf}. As can be seen it produces two beams from four elements. Each element for each beam is controlled for phase and amplitude with a 3-bit control word. The benefit of RF beamforming currently is that it is relatively cheap and low power compared to the digitization and processing required for digital beamforming.

\begin{figure*}
\centering
\includegraphics[width=0.6\linewidth]{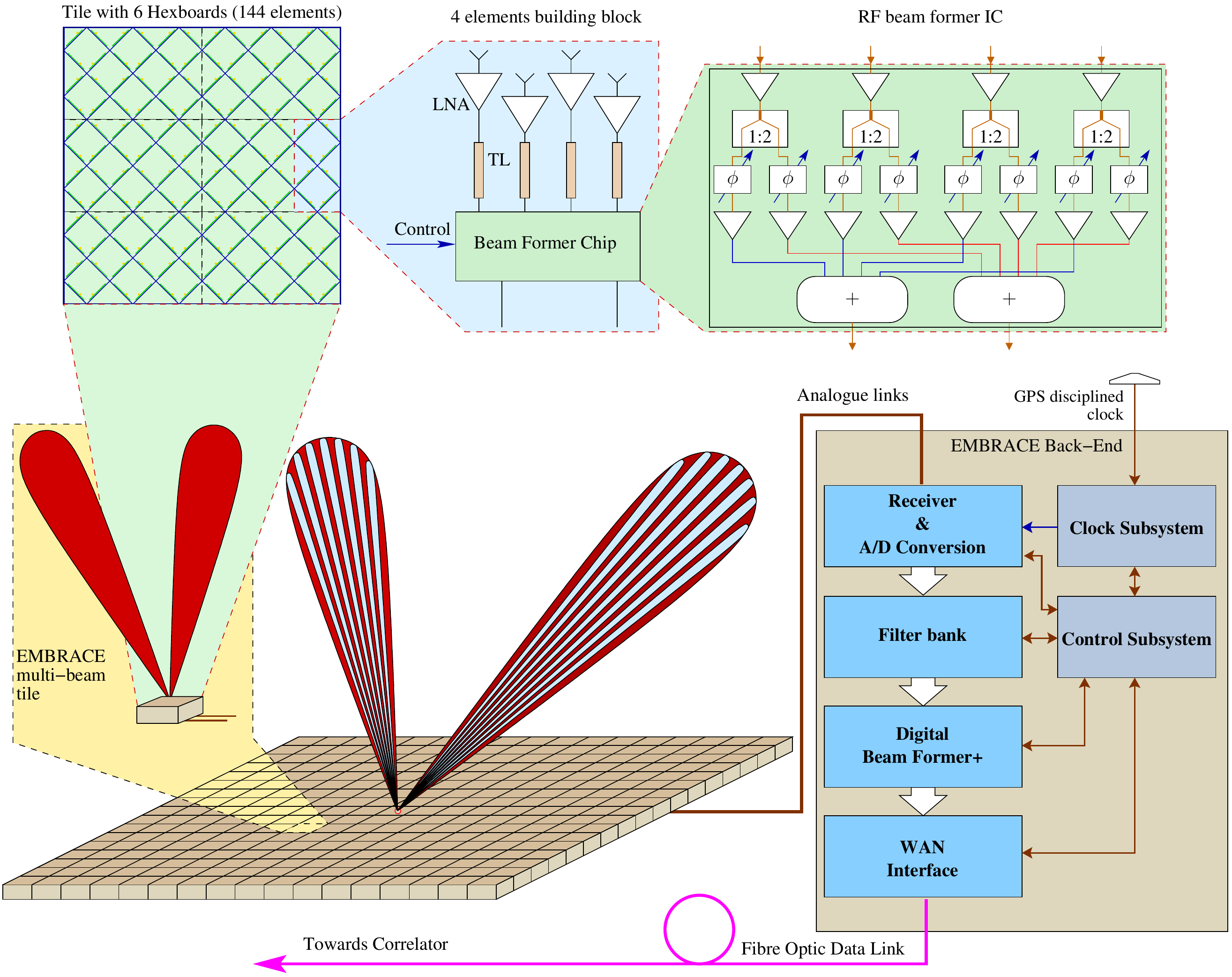}
\caption{System level overview of the EMBRACE station architecture. Antenna
  signals are combined at tile level using integrated circuit technology. Each
  tile provides two RF beam signals to a local processing facility where
  digital station beams are formed. Each tile uses two coaxial cables to
  transport the RF signals and also supply DC power and control signals to the
  tiles. This is reproduced from Kant et al. \ref{Kant}}
  \label{fig_embrace2_bf}
\end{figure*}

\subsection{EMBRACE Beamforming}

EMBRACE is an operational demonstrator developed to test the next generation radio telescopes. It demonstrates the possibility of independent, multiple, wide field, wide band receiving antenna beams based on phased array technology and described in detail in Monari et al. [\ref{Monari}]. EMBRACE operates in the 500--1500~MHz frequency range and consists of two aperture array stations implemented at Westerbork in the Netherlands and Nancay in France. Whilst the EMBRACE is inherently dual-polarisation the current implementation processes only one polarisation, with the second one being a trivial addition of twice as much hardware.

\begin{figure}[h!]
\begin{center}
\includegraphics[width=2.5in]{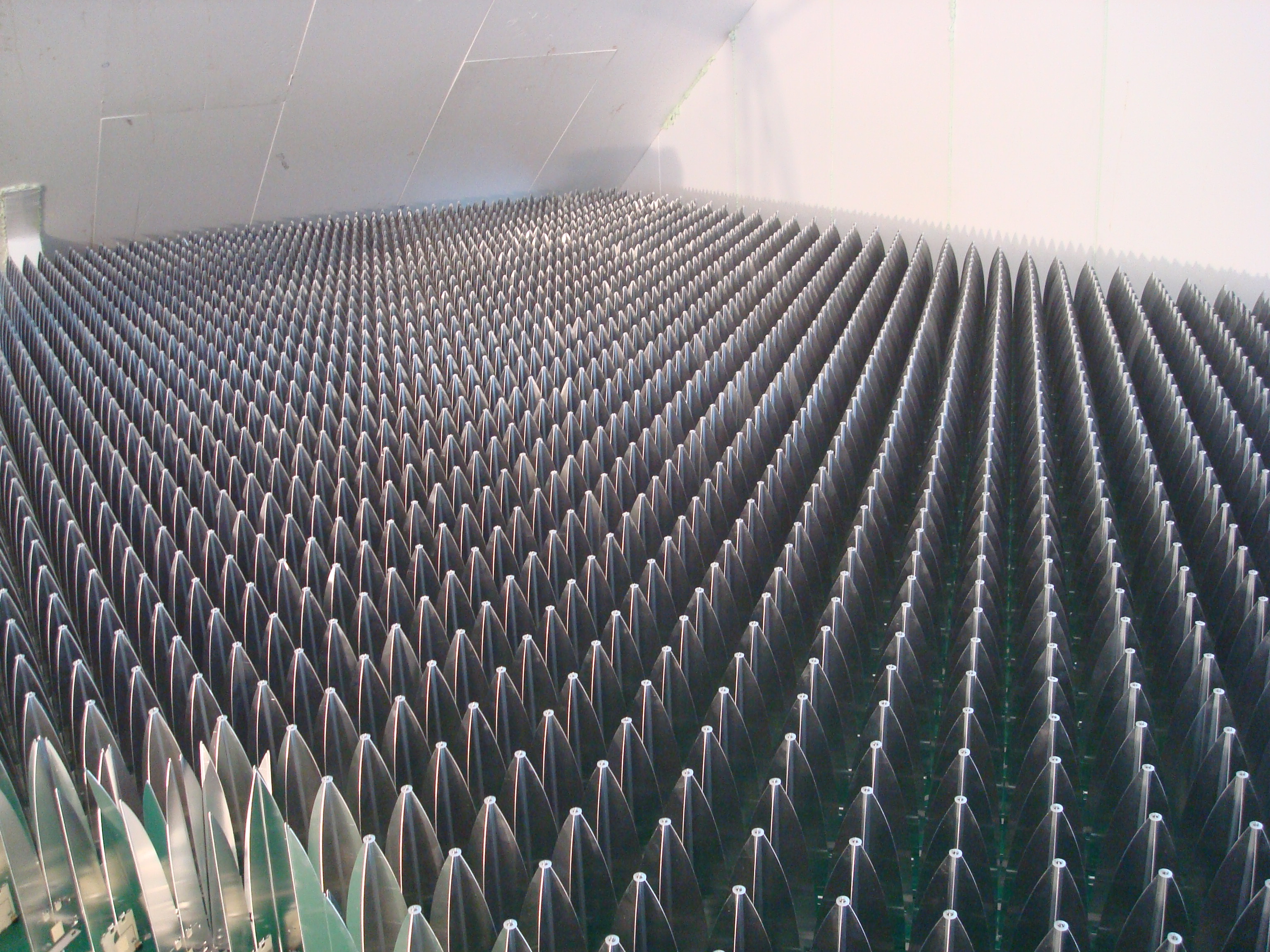}
 \caption{The EMBRACE Station.}
\label{fig_embrace3_bf}
\end{center}
\end{figure}

The base element of the EMBRACE arrays is the EMBRACE tile shown in figure \ref{fig_embrace3_bf}. Each tile has one control board with six identical `hex-modules' connecting 72 dual-polarised Vivaldi antennas. The tiles are then connected together to form a station according to the system diagram shown in figure \ref{fig_embrace2_bf}.

%\begin{figure}[!h]
%\includegraphics[width=5.0in]{embracesysv2.pdf}
% \caption{The EMBRACE System Diagram}
%\label{fig_embrace2_bf}
%\end{figure}

\subsection{RF Beam Former Chip}

An analogue RF MMIC has been developed, to control the phases of each element
in an EMBRACE tile. To generate phase shift, the design uses a vector
modulator approach, with a filter network network generating the four main
phase states 0, 90, 180 and 270 degrees. Smaller phase steps can be created by
combining these vectors. \\
\begin{figure}[!h]
\begin{center}
\includegraphics[width=2.0in]{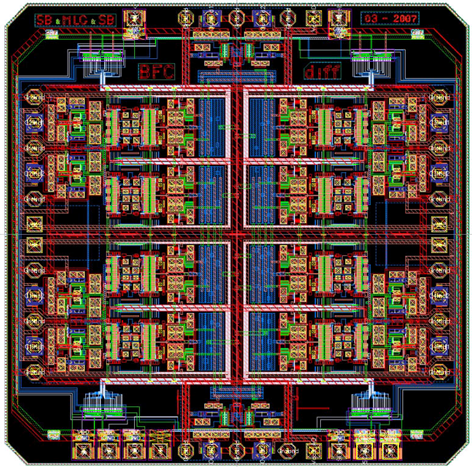}
 \caption{The EMBRACE RF-beamforming chip developed by OPAR. Chip photograph courtesy of P Picard.}
\label{fig_embrace_chip}
\end{center}
\end{figure}

The chip has been designed by OPAR, France in a QUBIC4G BiCMOS 0.25~$\mu$m
SiGe process of NXP and shown in figure \ref{fig_embrace_chip} and described in detail in Picard et al. [\ref{Picard}]. With this chip two times four differential channels are independently phase shifted, summed and amplified. Amplitude control is used to compensate for gain variation of the chip and the entire analogue signal transport chain. The phase settings are set by a digital serial control interface, to reduce layout complexity and crosstalk issues.

\section{Digital Beamforming}

With digital beamforming the incoming signal is firstly digitized, whether from a single element or the product of an initial RF beamformer. The signals from each channel are passed to a processing device for signal processing. This architecture is capable of very high performance in terms of bandwidth, precision, number of beams, and usage of the output data rate. Once the incoming signals are digitized then all the performance is at the cost of additional processing, memory and communications capacity. 

These parameters follow a growth law, either Moore's law or similar. It is likely therefore, that the role of digital beamforming will grow over time – eventually becoming cheaper in terms of power and cost than RF beamforming.

\begin{figure}[h!]
\centering
\includegraphics[width=70mm]{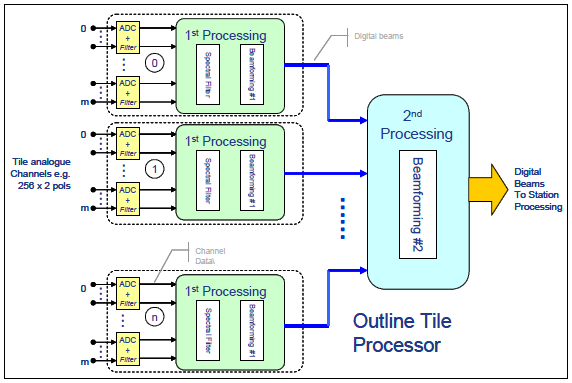}
\caption
	{An outline of a top-level digital beamforming system diagram} 
\label{outline_digital}
\end{figure}

Figure \ref{outline_digital} shows an outline of a signal processing system for the first stage of a digital beamformer. As noted, the first requirement is to digitize all the incoming channels. The system shown in the figure employs frequency domain beamforming which confers the maximum flexibility, so the signals are first passed through a polyphase filter bank, such that beamforming is then a matter of multiplying each spectral sample with a coefficient to provide phase rotation and amplitude adjustment, finally the sample is summed with samples from other elements to make a beam. This process is repeated to form as many beams as is required. Corrections can also be made for polarisation errors at this stage of processing, or at the tile level.

Time-domain digital beamforming is also possible and can be used to align signal paths at the sample level. However, when sub-sample alignment is required, interpolation filters need to be designed often adding to the complexity of the digital design. Arguments for broad-band digital time-domain beamforming are numerous and indeed valid, however in the case of the 2-PAD implementation a frequency domain architecture was selected since polyphase filter-banks allowed more control and flexibility on the correction of the analogue band-pass phase and amplitude characteristics.

\subsection{Beamforming on 2-PAD}

\begin{figure}[h!]
\centering
\includegraphics[width=70mm]{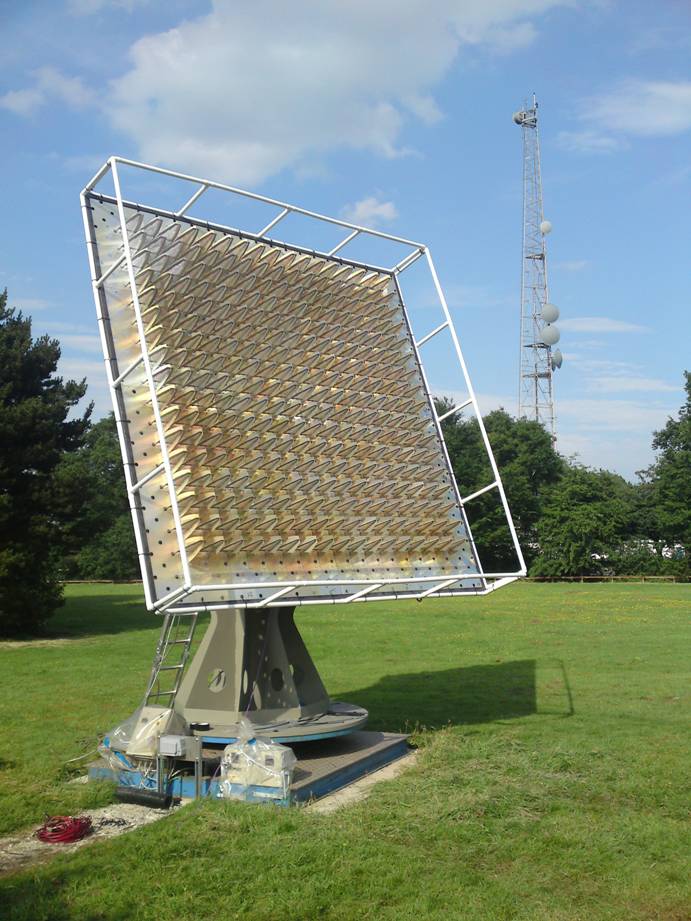}
\caption
	{The 2-Polarisation All-Digital (2-PAD) aperture array demonstrator at Jodrell Bank.} 
\label{fig_2-PAD}
\end{figure}

The 2-Polarisations All-Digital (2-PAD) project is a phased aperture array demonstrator built to confirm the feasibility of beamforming many, tightly-coupled signals with bandwidths of hundreds of Mega-hertz entirely in the digital domain and will demonstrate antenna design, RF conditioning and amplification, digitisation, dataflow and processing for future aperture array radio instruments\footnote{more information is available at \textsf{www.skads-eu.org}}. The 2-PAD demonstrator (shown in figure \ref{fig_2-PAD}) is a modular, end-to-end demonstrator system with a modest number of Bunny Ear Comb Antennas (BECA), described in Zhang et al. [\ref{Zhang}]. Due to its modular design, different implementations of subsystems (i.e. antennas, digitisers or processing systems) are able to be tested within a working system.

As discussed in the previous section, the broadband signal from each antenna is digitised using an eight-bit analogue-to-digital convertor (ADC) and then split into 1024 frequency bands. Each band is independently processed (delayed and combined with the signals from other antennas) with a narrowband phase-shift beamformer. The results of these narrowband beamformers are combined to form a broadband beam in the required spatial direction. This operation is repeated for each beam that is formed. Channelisation or `frequency binning' of large-bandwidth input signals for beamforming can be performed with various techniques, but a polyphase filter bank with an 8-tap FIR-filter and 10-stage FFT is implemented.

N single-polarisation antenna signals are digitised then combined with phase correction factors to form B beams in the beamformer processor. These beams are combined in a beam-combiner processor which could also be a further hierarchy of processors, combining more beams from other tiles.

\begin{figure}[h!]
\includegraphics[width=80mm]{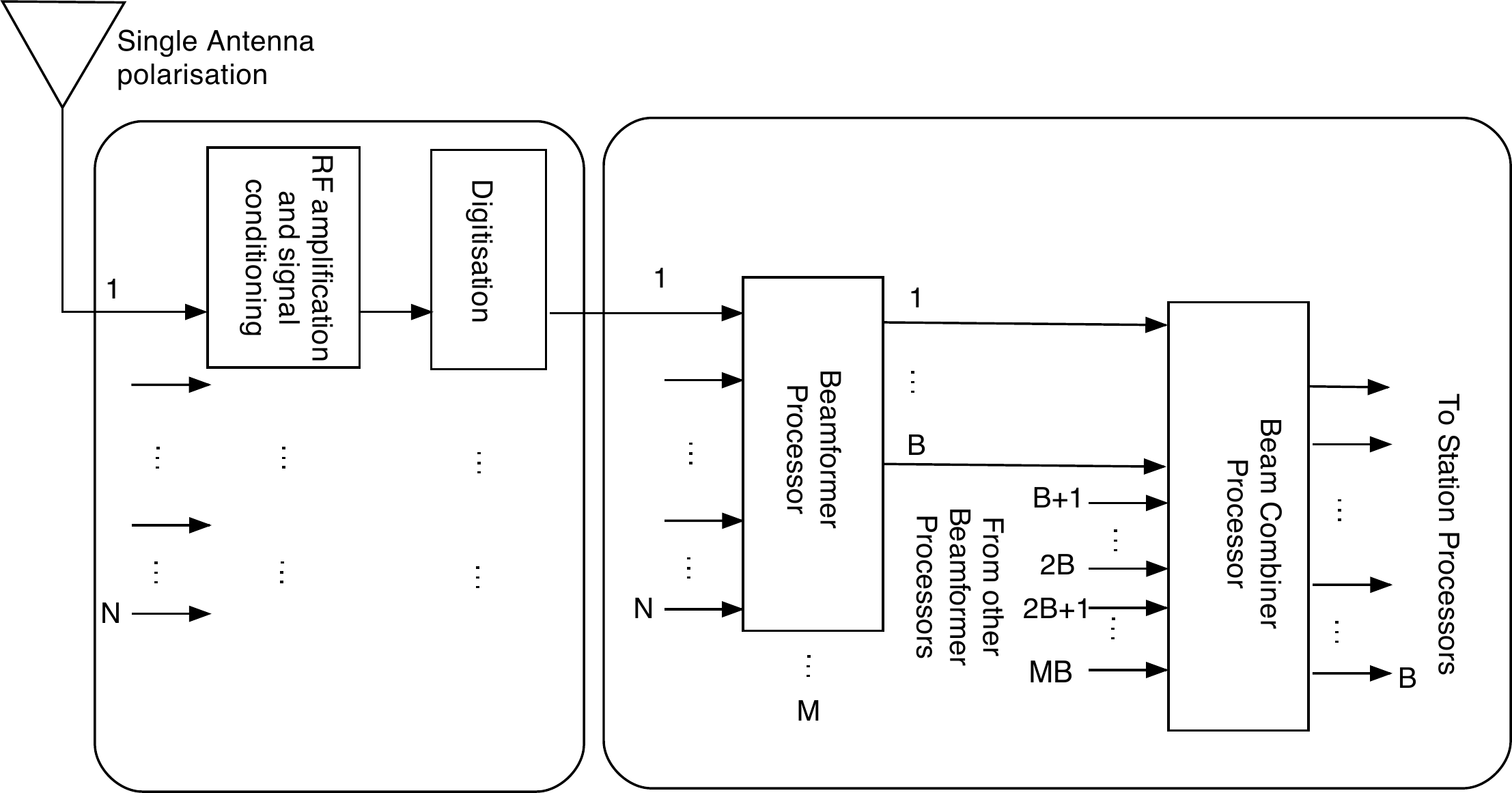}
\caption{System Signal Processing Flow for Digital Aperture Array. Only one polarisation is shown for clarity}
\label{SPFlowAA}
\end{figure}

For the 2-PAD case and in general for most digital beamforming systems being implemented, we find that rather than being processing limited we are limited by the I/O bandwidth and inter-processor bandwidth that is required to transport the signal from the antenna elements to the beam-forming processor. As an example in the 2-PAD system, the analogue input signal ranges from 0.5GHz to 0.7GHz (a bandwidth of 0.2GHz), and is sampled with 8-bits of precision, which results in a digital input bandwidth from each antenna polarisation of  3.2 Gbps. This means that the first-stage beamformer processor will have a raw input data-rate of $N \times 3.2$ Gbps and an output data rate of $B \times s \times 3.2$ Gbps, where $s$ is the bitwidth scaling factor of the beamformer.

Note that for slow moving objects, such as astronomical sources described in Thompson et al. \ref{Thompson}, there is a strong distinction between the high-performance, real-time processing which must be performed at the input data-rate and the more computationally complex beam-steering co-efficient calculation. In the next section, we will concentrate on the former; i.e an implementation of the high-performance signal processing system. We allow the steering and correction coefficients to be updated at a lower rate, dictated by the speed with which the array must be able to scan the sky.

\section{A Hierarchical Frequency Domain Beamforming Architecture}

Following from the previous section, we now describe the design and implementation of an FPGA-based digital system for the frequency sub-band phase-shift beamforming architecture.\\

\begin{figure}
\begin{center}
\includegraphics[width=70mm]{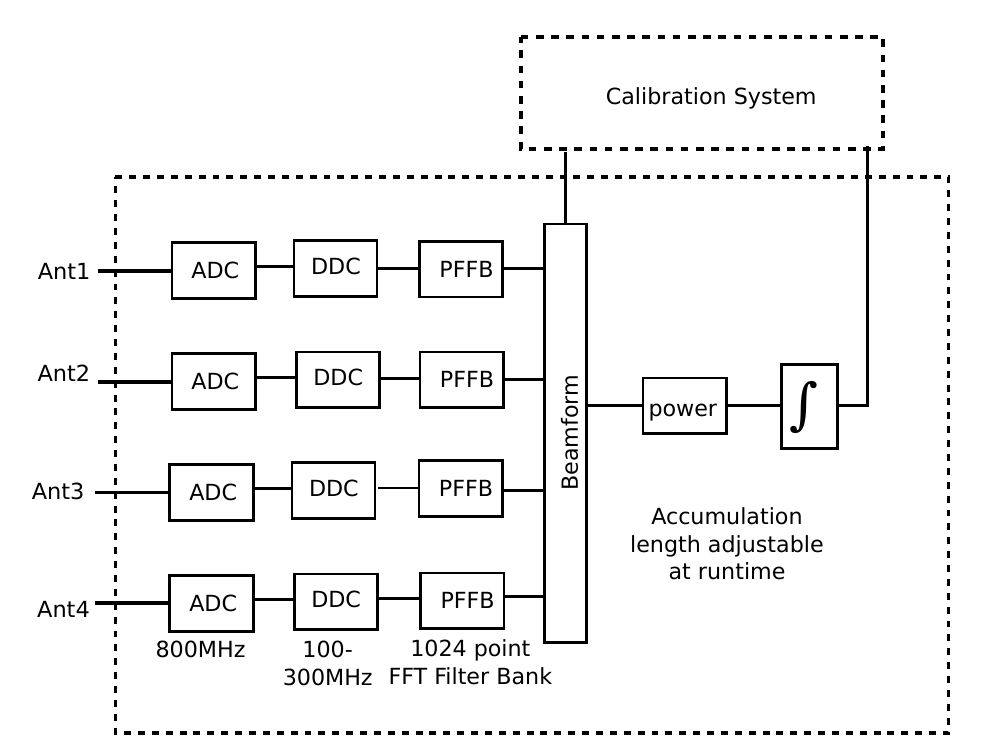}
\caption{An architectural drawing of the 4-element beamformer and calibration system} 
\label{2pad_arch}
\end{center}
\end{figure}

Figure \ref{2pad_arch} is an architectural schematic of a 4-element beamformer digital system. The Analogue to Digital Converter (ADC) samples the incoming antenna signal at 800MSa/s with 8 bits of precision. This 400MHz bandwidth signal is subsequently Digitally Down-Converted (DDC) by mixing with a complex sinusoid at 3/4 of the ADC clock rate to yield a base-banded 500MHz to 700MHz input signal. 

Channelisation is the next step in the processing pipeline: an efficient technique is the Polyphase FFT\footnote{Fast Fourier Transform (FFT)} Filter Bank (PFFB) channeliser, which improves frequency isolation of the FFT operation (see for example \ref{Armstrong}) and has been shown to be efficient for large numbers of equally spaced channels \ref{Armstrong}. The beamformer applies steering and correction coefficients to each signal stream and then performs beam summation. We then accumulate the power sum of the raw voltage beam signal for a run-time configurable length of time.

This design has been implemented using the signal processing libraries and hardware designed by the Collaboration for Astronomy Signal Processing Research (CASPER)\footnote{for more information, see \textsf{www.casper.berkeley.edu}}.

\begin{figure}
\begin{center}
\includegraphics[width=80mm]{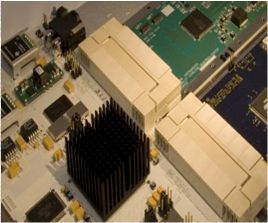}
\caption{A photo of the internet break-out board (iBOB) designed by CASPER} 
\label{ibob}
\end{center}
\end{figure}

\begin{figure}
\begin{center}
\includegraphics[width=80mm]{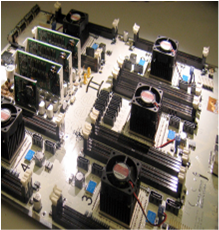}
\caption{A photo of the Berkeley Emulation Engine (BEE-2) board designed by CASPER} 
\label{bee2}
\end{center}
\end{figure}

The FPGA-based design is based on Xilinx Virtex II Pro devices which are available on the internet Break-Out-Boards (iBOB(s)) and the Berkeley Emulation Engine (BEE-2) boards shown in figure \ref{ibob} and \ref{bee2}. Since the FPGAs use fixed hardware blocks which are natively 18-bit, we preferentially use 18-bit multiplications and datapaths throughout the design.

\begin{figure}
\begin{center}
\includegraphics[width=70mm]{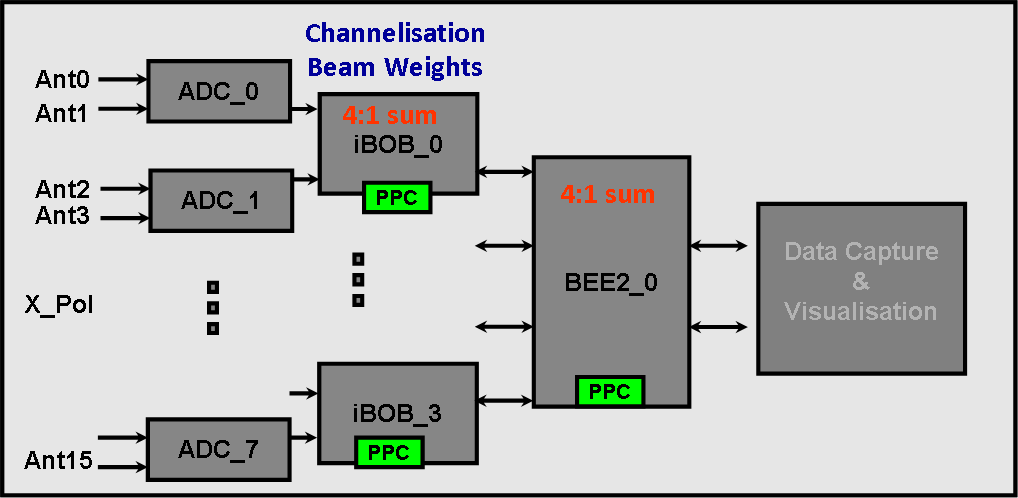}
\caption{An architectural drawing of a 16-element hierarchical beamformer} 
\label{hierarch_arch}
\end{center}
\end{figure}

A 16-element system based on a hierarchical architectures shown in figure \ref{hierarch_arch} has been implemented. Two simultaneous beams with an instantaneous bandwidth of 200MHz each were produced. Further information regarding the performance and calibration of these beams is presented in Armstrong et al. [\ref{Armstrong}].

\section{Discussion}

The bulk of the beamforming operation happens in the first stages both in terms of processing and bandwidth, since it is generally accepted that station level processing will be in a processor based solution due to the complexity and flexibility that is required. The question mostly revolves around how the initial stages of beamforming may be implemented. Here the consideration is essentially between RF beamforming and digital beamforming.  

Tables \ref{table_1} and \ref{table_2} summarises the trade-offs between RF and digital beamforming. It is clear that for significant arrays built in the short term that RF beamforming is an essential component, however, there are drawbacks in the capability and scaleability of the system. Digital beamforming through the whole array will confer substantial benefits and can provide the performance required by very large arrays but requires some reasonable and predictable semiconductor advances and significant development costs to implement.

\section{Conclusion}

The basic conclusion is that beamforming for large phased arrays is quite difficult to implement at low initial cost and in the short term, however, the application of significant development funds and the evolution of semiconductor technology makes the system not only feasible but economic. The discussion on the use analogue beamforming for the initial stages of the system compared to a complete system based on digital beamforming is going through a transition period. In the next ten years or so, for the frequencies in question here, a fully digital system will not only be much higher performance but also lower cost and power. This will only improve as longer term developments take place. The management of this period and the demonstration of performance using analogue techniques will be the challenge until then.

\section*{Acknowledgment}
Many people have contributed towards this review and the authors would especially like to thank Professor Mike Jones, Dr Paul Alexander, Professor Arnold van Ardenne, OPAR for very many discussions that led to the design and implemenation of the EMBRACE and 2-PAD systems.

This effort/activity is supported by the European Community Framework Programme 6, Square Kilometre Array Design Studies (SKADS), contract no 011938.

\ifCLASSOPTIONcaptionsoff
  \newpage
\fi

\newenvironment{figland}{
  \begin{sidewaysfigure}
  \small}{
  \end{sidewaysfigure}}

% tabland: sideways table without placement options
 
\newenvironment{tabland}{
  \begin{sidewaystable}
  \small}{
  \end{sidewaystable}}

\begin{landscape}
\begin{table}[!t]
%% increase table row spacing, adjust to taste
\renewcommand{\arraystretch}{1.3}
% if using array.sty, it might be a good idea to tweak the value of
%\extrarowheight as needed to properly center the text within the cells
\caption{RF and digital beamforming comparison}
\label{table_1}
\centering
%% Some packages, such as MDW tools, offer better commands for making tables
%% than the plain LaTeX2e tabular which is used here.
\begin{tabular*}{1\textwidth}{|p{30ex}|| p{60ex}| p{60ex}|}
\hline
Characteristic & RF Beamforming & Digital Beamforming \\
\hline\hline Implementation & Integrated into analogue chips with control interface. The inputs are multiple low level analogue signals, the devices may be mounted near to the LNAs. Each chip produces multiple beams from each block of input channels.& Each analogue signal needs to be amplified to levels acceptable to an ADC; these are digitized and passed to a processing device(s). The signals are split into multiple channels and beamformed in narrow channels.\\
\hline Beam Generation & Each beam is formed by phase shifting each input to provide the required delay and then summed for each chip. The amplitude may be adjusted for each input element independently. It is probably not practical to correct polarisation at the element level.
True time delay technology integrated onto chips is not currently proven, external delays become to large for practical implementation on a dense high frequency array.
RF beamforming chips may be cascaded for larger systems.
Each beam operates as a single frequency channel. This will restrict the number of tile beams that may be produced independently. & With the anticipated frequency domain beam forming, the beams are produced by phase shifting a narrow frequency band. Each channel may be calibrated for amplitude, or used for RFI excision. Polarisation may be corrected as a function of frequency. Each channel may be considered to be a sub-beam which can be used to construct beams with the required bandwidths.
Further beams may be produced by repeating the beamforming functions after spectral separation.
The output data rate determines the overall performance of the beamformer, assuming that there is sufficient processing available to produce the beams.
\\\hline
Multiple Beams & Each Tile beam needs to be produced via specific hardware within the beamformer chip. The configuration is fixed by the architecture design.& As discussed above the beams are made up of multiple sub-beams from specific frequency dependant coefficients. These can make up beams in any format required within the constraints of output data rates and processing.\\\hline
Bandwidth & Assuming that the beamformer is using phase shifting for time delays, or a frequency dependent time delay then the bandwidth will be restricted to some fraction of the operating frequency for each beam. Wider bandwidths can be constructed using multiple beams.
If true time delay can be produced then wider bandwidths up to the operational range of the elements and analogue system can be produced.& The digital system can operate over the full bandwidth available from the elements and analogue conditioning. This is because each of the sub-beams can be treated as a narrow independent beam.\\\hline
Bandpass Corrections & The bandpass corrections for each element need to be made in an overall fashion. It is unlikely that they can be adjusted for changing conditions. The corrections made will be identical for each beam. & The analogue system up to the ADC has to be flat enough for effective digitization to take place. Additional flexibility can be achieved through further digitisation resolution although this has cost and power implications.
The bandpass can be corrected as a function of frequency and if necessary by beam; each sub-band can be independently changed. \\\hline
Calibration &	The analogue beamformer can provide element level amplitude and approximate time delay calibration; neither of these are as a function of frequency. It is unlikely to be able to provide element level polarisation calibration since this is highly frequency and direction dependant.& The digital system can provide frequency and direction dependant calibration per beam. The calibration can be high resolution amplitude, phase and polarisation corrections for each sub-beam. Since many beams can be formed it is viable to dedicate a number of sub-beams to observe calibrated sources during observations to refine the calibration. \\\hline
Flexibility	& The characteristics of the RF beamformer are determined at build time for number of beams and bandwidth. The frequency band can be moved and the available beams can be independently steered.	& The digital beamformer is essentially flexible in all aspects up to the output data rate. So, numbers and bandwidth of beams, resolution, pointing etc are all flexible. \\\hline
\end{tabular*}
\end{table}
\end{landscape}

\newpage
%% Some packages, such as MDW tools, offer better commands for making tables
%% than the plain LaTeX2e tabular which is used here.
%\begin{tabular*}{1\textwidth}{|p{30ex}|| p{60ex}| p{60ex}|}
\begin{landscape}
\begin{table}[!t]
%% increase table row spacing, adjust to taste
\renewcommand{\arraystretch}{1.3}
% if using array.sty, it might be a good idea to tweak the value of
%\extrarowheight as needed to properly center the text within the cells
\caption{RF and digital beamforming comparison continued...}
\label{table_2}
\centering
%% Some packages, such as MDW tools, offer better commands for making tables
%% than the plain LaTeX2e tabular which is used here.
\begin{tabular*}{1\textwidth}{|p{30ex}|| p{60ex}| p{60ex}|}
\hline
Characteristic & RF Beamforming & Digital Beamforming \\\hline\hline
Power requirements &	This is a relatively simple system which should minimize the power requirements. Increasing the number of beams will increase power; hence the bandwidth-beam count product strongly affects the power needed. 
In the near term the power required for an implementable system will be minimized by using an amount of RF beamforming. It is to be determined what level of RF beamforming could be accepted and still meet SKA performance specification. &
The trend in power required will reduce over time, with greater integration  and particularly with a site with low RFI,	The power for a digital system depends initially on the analogue systems which will reduce over time and with integration, as with the RF beamforming. The digital system is strongly subject to improvements with: 
Silicon technology
Increased integration
Better processing architectures
Much of the above depends on the level of NRE spent, which must not be invested too early due to the continuing advances in technology.
There will be a time that they digital beamformer will be equivalent or lower power than an analogue beamformer; part of the ongoing work is to identify that transition time. \\\hline
Cost &	The basic cost of a simple analogue system is relatively low and currently substantially cheaper than a digital system. However, increasing performance has a high incremental cost associated with it, even assuming that it is practical to implement.
The costs will reduce with further advances in technology and with increasing integration, but only to a limit.	The current cost of a digital beamformer is relatively high, although for station beamforming it is affordable since so many elements are being processed.
The cost of a digital system will reduce dramatically over time for the reasons discussed under the power discussion. &
A digital beamformer has a relatively high basic cost since all the digitisers and spectral separation processing has to be implemented in order to make just one beam, however, the increment to add further performance is relatively low - there is only the need for relatively small amounts of additional processing and the associated communications. \\\hline
\end{tabular*}
\end{table}
\end{landscape}

% Note that IEEE does not put floats in the very first column - or typically
% anywhere on the first page for that matter. Also, in-text middle ("here")
% positioning is not used. Most IEEE journals use top floats exclusively.
% Note that, LaTeX2e, unlike IEEE journals, places footnotes above bottom
% floats. This can be corrected via the \fnbelowfloat command of the
% stfloats package.

% or



\end{document}